\documentclass[conference]{IEEEtran}
\IEEEoverridecommandlockouts
\usepackage{cite}
\usepackage{amsmath,amssymb,amsfonts}
\usepackage{graphicx,url,times,acronym,balance,booktabs}
\usepackage{textcomp}
\usepackage[table]{xcolor}
\usepackage{lipsum}

\def\BibTeX{{\rm B\kern-.05em{\sc i\kern-.025em b}\kern-.08em
		T\kern-.1667em\lower.7ex\hbox{E}\kern-.125emX}}

\acrodef{STFT}{short-time Fourier transform}
\acrodef{MSE}{mean-squared error}
\acrodef{MAE}{mean absolute error}
\acrodef{PSD}{power spectral density}
\acrodef{RTF}{relative transfer function}
\acrodef{SNR}{signal-to-noise ratio}
\acrodef{PDF}{probability density function}
\acrodef{DOA}{direction-of-arrival}
\acrodef{VAD}{voice activity detector}
\acrodef{MVDR}{minimum variance distortionless response}
\acrodef{PESQ}{perceptual evaluation of speech quality}
\acrodef{STOI}{short-time objective intelligibility}
\acrodef{LSD}{log spectral distance}
\acrodef{CD}{cepstral distance}
\acrodef{WER}{word error rate}
\acrodef{SPP}{speech presence probability}
\acrodef{DNN}{deep neural network}
\acrodef{RNN}{recurrent neural network}
\acrodef{CNN}{convolutional neural network}
\acrodef{FC}{fully connected}
\acrodef{CRN}{convolutional recurrent network}
\acrodef{LSTM}{long-term short-term memory}
\acrodef{GRU}{gated recurrent unit}
\acrodef{FF}{feed forward}
\acrodef{ReLU}{rectified linear unit}
\acrodef{PReLU}{parametric rectified linear unit}
\acrodef{GCC}{generalized cross-correlation}
\acrodef{RMSE}{root-mean-square error}
\acrodef{CPSD}{cross-power spectral density}
\acrodef{siSDR}{scale-invariant signal-to-distortion ratio}
\acrodef{SDR}{signal-to-distortion ratio}
\acrodef{LPS}{logarithmic power spectrum}
\acrodef{DL}{deep learning}
\acrodef{MOS}{mean opinion score}
\acrodef{RIR}{room impulse response}
\acrodef{AIR}{acoustic impulse response}
\acrodef{MAC}{multiply-accumulate}
\acrodef{ASR}{automatic speech recognition}
\acrodef{segVNR}{segmental voice-to-noise ratio}
\acrodef{VNR}{voice-to-noise ratio}
\acrodef{AUC}{area under curve}
\acrodef{BCE}{binary cross-entropy}

\definecolor{matlab1}{rgb}{0, 0.4470, 0.7410}
\definecolor{matlab2}{rgb}{0.8500, 0.3250, 0.0980} 
\definecolor{matlab3}{rgb}{0.9290, 0.6940, 0.1250} 
\definecolor{matlab4}{rgb}{0.4940, 0.1840, 0.5560} 
\definecolor{matlab5}{rgb}{0.4660, 0.6740, 0.1880}

\begin{document}
	
	\title{On training targets\\ for noise-robust voice activity detection}
	
	\author{\IEEEauthorblockN{1\textsuperscript{st} Sebastian Braun}
		\IEEEauthorblockA{\textit{Microsoft Research} \\
			Redmond, WA, USA \\
			sebastian.braun@microsoft.com}
		\and
		\IEEEauthorblockN{2\textsuperscript{nd} Ivan Tashev}
		\IEEEauthorblockA{\textit{Microsoft Research} \\
			Redmond, WA, USA \\
			ivantash@microsoft.com}
	}
	
	\maketitle
	
	\begin{abstract}
		The task of voice activity detection (VAD) is an often required module in various speech processing, analysis and classification tasks. While state-of-the-art neural network based VADs can achieve great results, they often exceed computational budgets and real-time operating requirements. In this work, we propose a computationally efficient real-time VAD network that achieves state-of-the-art results on several public real recording datasets. We investigate different training targets for the VAD and show that using the segmental voice-to-noise ratio (VNR) is a better and more noise-robust training target than the clean speech level based VAD. We also show that multi-target training improves the performance further.
	\end{abstract}
	
	
	\section{Introduction}
	Classifiers detecting speech presence in audio signals, widely known as a \ac{VAD}, are a mature but still popular research topic. There are countless applications that require or benefit from a \ac{VAD}, like all kinds of audio processing modules \cite{Hansler2004,Tashev2009} like speech enhancers, localizers, echo controllers, automatic gain controls, speech level or \ac{SNR} measurement, speech-rate or silence-rate estimation, pre-processing, gating or segmentation for speech recognizers, speech-related classifiers for emotion \cite{Schuller2011}, gender, age, identity, anomaly, toxicity, or speech encoding and transmission \cite{Baeckstroem2017}, and so forth.
	
	Traditional statistical models, widely used in speech enhancement due to their simplicity, often rely only on the non-stationarity of speech \cite{Sohn1999,Gerkmann2012,Martin2001}. Better accuracy can be achieved by integrating models of features like pitch, harmonicity \cite{Dhananjaya2010,Tan2010}, modulation \cite{Hsu2015}, spectral shape \cite{Sadjadi2013,Yoo2015}, etc. \cite{Graf2015}. Dealing with increasing numbers of features becomes more difficult with human-crafted models, often resulting in diminishing performance gains. Therefore, data-driven approaches, especially neural networks \cite{Eyben2013,Hughes2013,Vesperini2016}, are an attractive choice and have shown substantial performance boosts \cite{Tashev2016}.
	
	A major design point of \ac{VAD} methods is the temporal resolution of the classification. While for some applications like speech segmentation, emotion classifiers etc.\ a larger granularity is sufficient, speech processing applications usually require a \ac{VAD} in frame-level resolution of typical lengths between 5 - 20~ms. While some \ac{DL} based \acp{VAD} designed for the first class of applications use coarser temporal resolutions of 0.2 - 10~s, which additionally can improve robustness and performance, this limits the use as general purpose \ac{VAD} for the second class of applications. Furthermore, most applications have real-time requirements, and are deployed on computationally and power-constrained edge devices, which poses challenges on the latency and computational efficiency of the algorithms.
	
	In this work, we propose a neural network architecture for real-time \ac{VAD} on a typical short audio frame basis. The network provides predictions per frame without look-ahead, and is reasonably small and efficient to be executed on standard CPUs of typical battery-powered devices. The main contribution of this work is, however, the investigation of noise robust \ac{VAD} training targets. We show that training on the clean speech level-based \ac{VAD} is less robust in low \ac{SNR} conditions than using the frequency-weighted segmental \ac{VNR}. While the ground truth depending only on the clean speech level is ill-conditioned in very low \ac{SNR}, where the speech might not even be audible, the \ac{VNR} label accounts for the audibility of speech in noise. Additionally, we propose a multi-target training, using speech level-based \ac{VAD} and \ac{VNR} as joint targets, which improves the performance further.
	While in \cite{Li2020a} a method for frame-level \ac{SNR} estimation has been proposed, to the best of our knowledge this is the first work to use \ac{SNR} as training target for \ac{VAD}.


	\section{Signal model and training targets}
	\label{sec:sigmodel}
	We assume that noisy training data is generated by mixing possibly reverberant speech with noise, resulting in the training mixture signal
	\begin{equation}
		\label{eq:signal_model}
		y(t) = h(t) \star s(t) + v(t),
	\end{equation}
	where $h(t), s(t), v(t)$ are the \ac{AIR}, non-reverberant speech signal and noise signal, $t$ is the time index, and $\star$ denotes the convolution operator. As we are interested in detecting speech, be it reverberant or not, but do not consider reverberant tails as desired speech information, we define the target speech signal 
	\begin{equation}
		\label{eq:target_speech}
		x(t) = h_\text{win}(t) \star s(t),
	\end{equation}
	where $h_\text{win}(t)$ is a windowed version of the full \ac{AIR} $h(t)$ removing the late reverberation tail. We chose an exponentially decaying window corresponding to a decay rate of $60~\text{dB} / 0.3~s$ starting from the direct path of the \ac{AIR}.

	The typical \ac{VAD} training target is defined by the clean speech level threshold $T_\text{level}$ as
	\begin{equation}
		\label{eq:vad}
		VAD(n) = \begin{cases} 1 \quad\text{if}\; \Vert\mathbf{W}_\text{VAD}\, \mathbf{x}(n)\Vert^2 > T_\text{level}\\ 
			0 \quad\text{if}\; \Vert\mathbf{W}_\text{VAD}\, \mathbf{x}(n)\Vert^2 \leq T_\text{level}\end{cases},
	\end{equation}
	where the vector $\mathbf{x}(n)$ contains the frequency-domain representation of the target speech signal $x(t)$ at frame $n$ and the diagonal matrix $\mathbf{W}_\text{VAD}$ is an optional frequency weighting such as bandpass filters or loudness weighting.
	
	We propose an alternative training target, i.\,e.\ the Mel-weighted segmental \ac{VNR}
	\begin{equation}
		\label{eq:segVNR}
		VNR(n) = 10\log_{10} \frac{\Vert\mathbf{W}_\text{VNR}\, \mathbf{x}(n)\Vert^2}{\Vert\mathbf{W}_\text{VNR}\, \mathbf{v}(n)\Vert^2},
	\end{equation}
	where $\mathbf{v}(n)$ is the frequency-domain representation of the noise signal $v(t)$ and the matrix $\mathbf{W}_\text{VNR}$ is an auditory or loudness weighting, e.\,g.\ a Mel-filterbank.
	Speech presence can then be computed similarly as in \eqref{eq:vad} by comparing the continuous VNR values to a threshold. Note that in contrast to the speech level only dependent \ac{VAD}, the VNR also takes noise into account, providing additional information to voice presence about the background noise and \emph{the audibility of the speech in noise}. This avoids practically wrong \ac{VAD} labels \eqref{eq:vad} for highly negative \ac{SNR}, i.\,e. when the speech may not even be audible. Furthermore, the VNR attributes a physical and interpretable meaning to the speech detection threshold. By adjusting the detection threshold for a VNR based \ac{VAD}, we can detect either any audible speech in noise using a very low threshold, or only target well-audible foreground speech by choosing a positive VNR threshold. The term \ac{VNR} is used in analogy to \ac{VAD}, to highlight measurement of voice only, opposed to the general \ac{SNR}.
	

	\section{Proposed single- and multi-target losses}
	\label{sec:losses}
	Having defined the two training targets, the binary $VAD(n)$, and the continuous $VNR(n)$ in the previous section, we define the considered loss functions in the following. A natural choice for a binary classification like $VAD(n)$ is the \ac{BCE} \cite[ch. 2.8.1]{Murphy2012}
	\begin{equation}
		\label{eq:bce}
		\mathcal{L}_\text{BCE}(z) = - \frac{1}{N} \sum_n z_n \log(\hat{z}_n) + (1-\hat{z}_n) \log(z_n),
	\end{equation}
	where $z_n$ and $\hat{z}_n$ are labels and predictions, respectively, and $N$ is the number of frames.
	On the other hand, continuous distributions like $VNR(n)$ are often fitted using the \ac{MSE} or \ac{MAE} \cite{Li2020a}. We indeed determined in preliminary experiments that the \ac{BCE} performed best when training $VAD(n)$, while the \ac{MAE} loss performed best when training to predict $VNR(n)$.
	
	As multi-target training can be conveniently realized in deep learning and often helps generalization and robustness, we propose to combine $VAD(n)$ and $VNR(n)$ training targets. We explored applying \ac{BCE} and \ac{MAE} loss to the training targets, resulting in the two multi-target training loss options
	\begin{align}
		\mathcal{L}_\text{VADVNR,1} &= (1\!-\!\alpha) \mathcal{L}_\text{BCE}(VAD) + \alpha \mathcal{L}_\text{MAE}(VNR) \label{eq:vadvnr_bcemae}\\
		\mathcal{L}_\text{VADVNR,2} &= \mathcal{L}_\text{BCE}(VAD) + \mathcal{L}_\text{BCE}(VNR), \label{eq:vadvnr_bce}
	\end{align}
	where the weighting factor $\alpha=0.2$ to balance \ac{BCE} and \ac{MAE} was optimized on the validation set.
	We finally consider four different loss functions:
	\begin{enumerate}
		\item Predict VAD with \ac{BCE} loss: $\mathcal{L}_\text{BCE}(VAD)$
		\item Predict VNR with \ac{MAE} loss: $\mathcal{L}_\text{MAE}(VNR)$
		\item Predict both VAD, VNR with \ac{BCE} and \ac{MAE}: $\mathcal{L}_\text{VADVNR,1}$
		\item Predict both VAD and VNR with \ac{BCE}: $\mathcal{L}_\text{VADVNR,2}$
	\end{enumerate}

	\section{Proposed network architecture}
	Due to the success of \ac{CRN} structures for efficient noise suppression \cite{Tan2018,Braun2021a}, we adapt a similar structure for the \ac{VAD} classification task. 
	\begin{figure}[tb]
		\centering
		\includegraphics[width=\columnwidth,clip,trim=250 195 180 120]{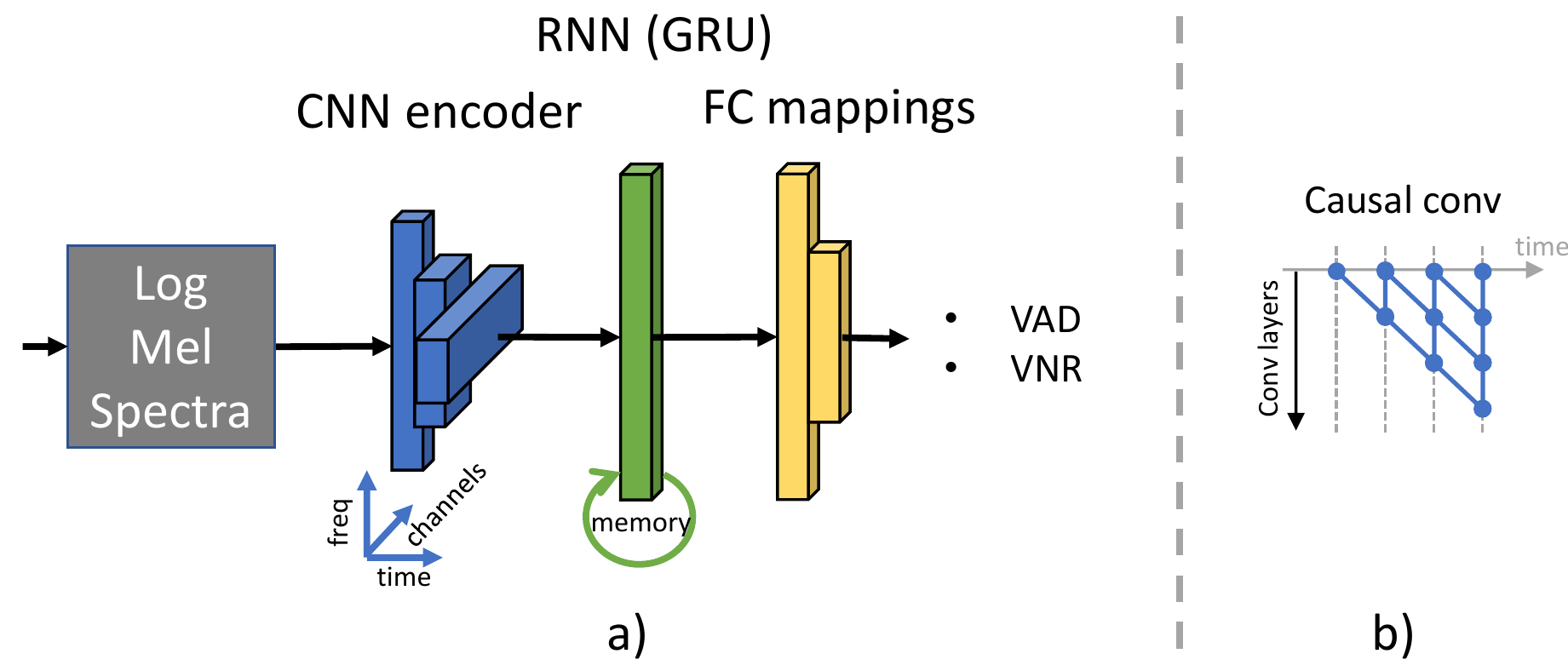}
		\caption{a) convolutional recurrent network architecture, b) efficient causal convolutions with temporal kernel size 2.}
		\label{fig:architecture}
	\end{figure}
	Figure~\ref{fig:architecture}a) shows the proposed network architecture. To reduce the input dimensionality, the input features are log-Mel energies. The features are encoded by a few 2D convolutional layers extracting spectro-temporal information. The frequency axis is reduced by a stride of 2 every layer, while the channel dimension is doubled. The convolution over time is causal, meaning no future information is used to infer the current frame as illustrated in Fig.~\ref{fig:architecture}b). The kernel size along time axis is only 2, which is very efficient, but still extracts temporal information across $N_\text{CNN}+1$ frames, where $N_\text{CNN}$ is the number of convolution layers.
	The output of the convolutional encoder is reshaped to a single vector and fed to a recurrent layer, specifically a \ac{GRU} \cite{Cho2014}. Finally, two \ac{FC} layers are used to obtain the desired output of one (single-target case) or two values (multi-target case) per time frame. All convolutional and \ac{FC} hidden layers use \ac{PReLU} activations, while the output layer uses a sigmoid to obtain constrained values.

	\section{Experimental validation}
	
	\subsection{Training data}
	We a use large-scale augmented synthetic training set to ensure generalization to real-world signals. The training set uses 544~h of high \ac{MOS} rated speech recordings from the LibriVox corpus, 247~h noise recordings from Audioset, Freesound, internal noise recordings and 1~h of colored stationary noise. Except for the 65~h internal noise recordings, the data is available publicly as part of the 2nd DNS challenge\footnote{https://github.com/microsoft/DNS-Challenge} \cite{Reddy2021}. 
	Non-reverberant speech files were augmented with acoustic impulse responses randomly drawn from a set of 7000 measured and simulated responses from several public and internal databases. 20\% non-reverberant speech is not reverb augmented to represent conditions such as close-talk microphones or professionally recorded speech. 
	
	Time shifted speech and noise sequences were mixed with a Gaussian \ac{SNR} distribution with $\mathcal{N}(5, 10)$~dB and augmented to different microphone signal levels with $\mathcal{N}(-28, 10)$~dBFS. The total training set comprised roughly 1000~h of 10~s mixture-target signal pairs. A more detailed description of the dataset generation can be found in \cite{Braun2021a}. The training targets $VAD(n)$ and $VNR(n)$ are obtained as described in Section~\ref{sec:sigmodel}.
	
	For training monitoring and hyper-parameter tuning, we generated a synthetic validation set in the same way as above, using speech from the DAPS dataset \cite{Mysore2015}, and \acp{RIR} and noise from the QUT\footnote{https://research.qut.edu.au/saivt/databases/qut-noise-databases-and-protocols/} database.
	
	\subsection{Experimental setup}
	Training targets and features were computed from 32~ms frames with 16~ms frame shift. The VAD weighting $\mathbf{W}_\text{VAD}$ was a bandpass filter between $[150,5000]$~Hz, and we used a signal-dependent threshold $T_\text{level} = 0.01 \cdot \max \left[ |\mathbf{W}_\text{VAD}\, \mathbf{x}(n)|^2\right]$. 
	$\mathbf{W}_\text{VNR}$ was a 32 band Mel weighting, and $VNR(n)$ was limited to the range $[-15,40]$~dB and mapped to the range $[0,1]$ for training purposes. Both targets, $VAD(n)$ and $VNR(n)$ were temporally smoothed to remove measurement errors and obtain smoother predictions, using a centered moving average window smoothing with length 0.2~s. 

	The network input features were 64 log-Mel energy bins in the range 0-8~kHz. 
	\begin{table}[tb]
		\centering
		\caption{Neural network layer dimensions.}
		\label{tab:parameters}
		\begin{tabular}{l|cc}
			\toprule
			layer type & hyperparameters & activation \\
			\midrule
			conv2D & 1 $\rightarrow$ 16, (2,3), (1,2), (1,0,1,1) & PReLU \\
			conv2D & 16 $\rightarrow$ 32, (2,3), (1,2), (1,0,1,1) & PReLU \\
			conv2D & 32 $\rightarrow$ 64, (2,3), (1,2), (1,0,1,1) & PReLU \\
			conv2D & 64 $\rightarrow$ 128, (2,3), (1,2), (1,0,1,1) & PReLU \\
			reshape	& 128$\times$4 $\rightarrow$ 512 & -- \\
			GRU & 512 $\rightarrow$ 512 & Sigmoid, tanh \\
			FC & 512 $\rightarrow$ 256 & PReLU \\
			FC & 256 $\rightarrow$ 1/2 & Sigmoid \\
			\bottomrule
		\end{tabular}
	\end{table}
	Table~\ref{tab:parameters} shows the network layer dimensions. Convolutional layers have parameters \emph{\{inChannels $\rightarrow$ outChannels, kernelSize, stride, padding\}}, where \emph{kernelSize} and \emph{stride} are defined as \emph{(time,frequency)}, and \emph{padding} is defined as \emph{(timeLeft,timeRight,freqLow,freqHigh)}. Reshaping, uni-directional GRU, and \ac{FC} layers are defined by \emph{\{inShape $\rightarrow$ outShape\}}.
	The proposed networks are trained using the AdamW optimizer \cite{Loshchilov2019} with a learning rate of $5\cdot 10^{-5}$, weight decay 0.01, batch size of 50 sequences of 10~s length, and adaptive gradient norm clipping \cite{Seetharaman2020}. Training is stopped when there is no improvement anymore on the validation set.

	\subsection{Metrics, test sets, and post-processing for evaluation}
	We used the \ac{AUC} of false positive rate vs. true positive rate as metric for evaluation, training monitoring and tuning. In contrast to the often used precision and recall or F-score, \ac{AUC} is threshold-independent, and we believe it is therefore more holistic.
	
	As \ac{DL} methods can only be convincing when generalization on real data can be proven, we use three different public datasets consisting only of real recordings in the wild. We chose the datasets to be from public domains, having been
	used in prior studies to benchmark \ac{VAD} methods for comparability, and to be representative for common real-world scenarios and challenging \ac{SNR} distributions. The three test sets are described in the following. The \emph{KAIST} dataset\footnote{https://github.com/jtkim-kaist/VAD} consists of four 30-min field-recordings in noisy environments (bus stop, construction site, park, room) performed by the authors of \cite{Kim2018}. Speech on- and off-sets are human-labelled by the authors. The \emph{HAVIC} database \cite{Strassel2012} is a collection of 72~h videos of everyday life situations like children playing, kitchens, living rooms, traffic, office, sports events, etc., with human annotations of acoustic events. We categorized all speech-related events as target, while the labels \{noise, background noise, music, unintelligible, baby, TV, singing\} were considered as non-speech. Although baby and singing are voice, it was excluded as it is not represented in our speech training data. The \emph{AVA speech v1.0} dataset\footnote{http://research.google.com/ava/index.html} is a 30~h collection of crowd-source labeled YouTube clips (we were only able to obtain 120 of the 160 15-min clips from YouTube). Most of the AVA speech clips seem to originate from movies, so this dataset should be considered as a biased subsample of possible scenarios.
	
	As all test sets are human-labeled, this might involve annotation errors, and exhibit coarse temporal granularity (pauses between words and sentences are usually still labeled as speech), and temporal label uncertainty. This might interfere with our fine-grained frame-wise \ac{VAD} predictions, that will detect short pauses between words.
	To mitigate these test label inaccuracies, we apply a post-processing to the frame-wise predictions $VAD(n)$, $VNR(n)$ by smoothing the predictions with a fixed window of 0.4~s length. The window involves no look-ahead, computing the 90-th percentile within the window, in order to detect speech, even when only a fraction of the frame within the 0.4~s window contains speech activity. This post-processing achieved up to 6\% relative AUC improvement on the given datasets, while still being real-time compatible.

	\subsection{Results}
	We compare our models to a state-of-the-art baseline ACAM \cite{Kim2018} employing a recursive temporal attention module and features with look-ahead of 190~ms.
	Table~\ref{tab:results} shows the \ac{AUC} results for baseline and proposed models. We can see that our best model achieves comparable performance on KAIST and better performance on the diverse and noisy HAVIC dataset, while being real-time without look-ahead, and using a less complex model during inference time. The ONNX model of our network processes 1~s audio in 7~ms on a laptop CPU at 2.5~GHz, while for ACAM \cite{Kim2018} reported 10~ms without, and over 200~ms with feature extraction on a GPU workstation.
	
	\begin{table}[tb]
		\centering
		\caption{AUC results (\%) on test sets.}
		\label{tab:results}
		\begin{tabular}{lll|cccc}
			\toprule
			target & loss & output		& KAIST	& HAVIC	& AVA\\
			\midrule
			\rowcolor{gray!25}
			ACAM \cite{Kim2018} & &			& \bf{99.2}	& 73.4	& n.\,a.\ \\	
			VAD \eqref{eq:vad} & BCE \eqref{eq:bce} & VAD					& 95.9	& 86.8	& 92.4\\
			\rowcolor{gray!25}
			VNR \eqref{eq:segVNR} & MAE & VNR					& 98.8	& 88.1	& 92.3\\
			VAD,VNR & BCE,MAE \eqref{eq:vadvnr_bcemae} & VNR 		& 98.8	& \bf{88.3}	& \bf{92.6}\\
			VAD,VNR & BCE,MAE \eqref{eq:vadvnr_bcemae} & VAD 		& 95.9	& 86.5	& 92.0\\
			\rowcolor{gray!25}
			VAD,VNR & BCE \eqref{eq:vadvnr_bce} & VNR		 	& \bf{99.0}	& \bf{88.9}	& \bf{92.4}\\
			\rowcolor{gray!25}
			VAD,VNR & BCE \eqref{eq:vadvnr_bce} & VAD			& 93.9	& 86.1	& 91.4\\
			\bottomrule
		\end{tabular}
	\end{table}
	The first and second colums in Table~\ref{tab:results} indicates the training targets and loss functions of the proposed networks. As described in Section~\ref{sec:losses}, we trained four different networks, two single-target and two multi-target predictors. Since the multi-target networks provide two outputs per frame, they are evaluated in two ways, either using the VAD or the VNR output.
	While the multi-target training improves performance on the test sets, we did not find useful improvements by combining the VAD and VNR outputs of the model, compared to only using VNR output.
	The two top results per dataset are printed boldface. We can observe that predicting VNR yields significantly better results than VAD on KAIST and HAVIC, while the difference on AVA is minor. The VNR output of the multi-target predictors show consistently equal or better results than the single-target predictors, while the VAD output of the multi-target networks performs still worse than the VNR single target. This illustrates both the advantage predicting VNR over VAD, and also the slight gain by multi-target training.
	Note that the label errors of the HAVIC dataset prevent achieving high results.
	
	\begin{figure}[tb]
		\centering
		\includegraphics[width=0.95\columnwidth,clip,trim=0 5 0 18]{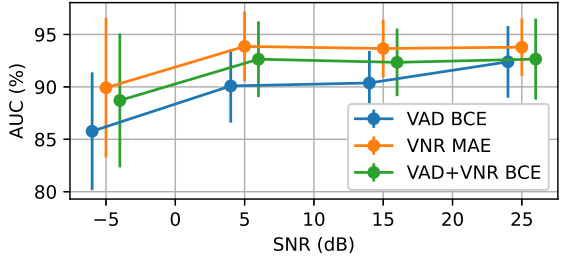}
		\caption{AUC (mean, std.\ dev.) on validation set for training targets over SNR.}
		\label{fig:auc_snr}
	\end{figure}
	Further analysis of AUC on the validation set grouped by the mixing \ac{SNR} shown in Fig.~\ref{fig:auc_snr} reveals that indeed all training targets perform similarly at high \ac{SNR}, while VNR and VAD+VNR targets provide better performance at medium to low \ac{SNR}. This proves the robustness to noise of using \ac{VNR} as training target compared to the common clean speech level based activity.
	Interestingly, on the validation set the VNR only (orange) loss shows slightly better results than the multi-target loss (green line) in Fig.~\ref{fig:auc_snr}, while all three test sets in Table~\ref{tab:results} show a slight advantage of the multi-target training. We still can conclude that the multi-target loss helps generalization due to better results on three real recording test sets in contrast to the synthetically mixed validation set, which the networks also were optimized on, and is therefore indirectly seen data.

	\begin{figure}[tb]
		\centering
		\includegraphics[width=\columnwidth,clip,trim=30 0 50 20]{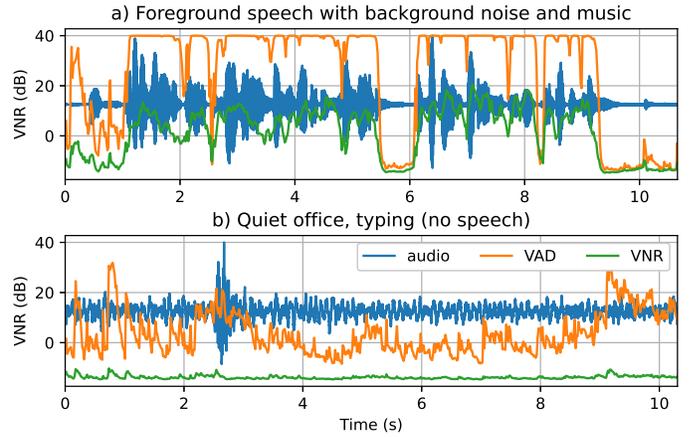}
		\caption{Outputs of the proposed model trained to output VAD and VNR with BCE loss for a) an easy scenario with prominent foreground speech, b) a recording without active speech.}
		\label{fig:output}
	\end{figure}
	The two outputs of the best proposed model, multi-target prediction with BCE loss, is shown in Fig.~\ref{fig:output} for two test recordings. The waveform is shown in blue, the VAD output in orange, and the VNR in green. Waveform and VAD (range [0,1]) are scaled to the VNR dB y-axis for illustration purposes. For the easy scenario in Fig.~\ref{fig:output}a), both outputs provide good indications of active and non-active speech frames. Both predictors show low values even in short pauses between words. However, a recording in a quiet office, containing only slight ambient noise and keyboard typing in Fig.~\ref{fig:output}b) reveals some problems of the classification VAD output. The output is more noisy, and the VAD predictions become quite large for some non-speech acoustic events. On the other hand, the VNR prediction is consistently low, hovering around -10~dB, indicating very unlikely speech activity at any time. It is worthwhile mentioning that the lowest equal error rate (false alarm equals miss rate) on the validation set is achieved with a \ac{VNR} threshold of -7~dB.

	\section{Conclusions}
	We have proposed an efficient real-time neural network based VAD that achieves state-of-the-art results on challenging real recordings. We showed that the segmental \ac{VNR} is a more noise robust training target for \ac{VAD} than the clean speech level based activity, and can be further improved by combining both targets for multi-target training. The proposed network is flexible for most applications, as the frame-level based decisions can be converted to coarser granularity by simple post-processing. The efficient network design allows straightforward integration in various speech processing tasks and different implementation platforms, as inference time on a modern CPU is around 7~ms per second of audio without any runtime optimizations.

	
	\pagebreak
	\balance
	
	\bibliographystyle{IEEEtran}
	\bibliography{sapref.bib}

\end{document}